\documentstyle[prb,twocolumn,epsf,aps]{revtex}

\begin{document}
\draft
%
%
\newcommand{\yn}{YNi$_{2}$B$_{2}$C}
\newcommand{\ynp}{Y(Ni$_{1-x}$Pt$_{x}$)$_{2}$B$_{2}$C}
\newcommand{\tc}{$T_{c}$}
\newcommand{\msr}{$\mu$SR}

\twocolumn[\hsize\textwidth\columnwidth\hsize\csname@twocolumnfalse\endcsname

\title{Nonlocal Effects and Shrinkage of the Vortex Core Radius in \yn\ Probed by \msr}
\author{K. Ohishi, K. Kakuta and J. Akimitsu}
\address{Department of Physics, Aoyama-Gakuin University, Setagaya-ku, Tokyo 157-8572, Japan}
\author{W. Higemoto and R. Kadono\cite{soken}}
\address{Institute of Materials Structure Science, High Energy Accelerator Research Organization (KEK), \\
Tukuba, Ibaraki 305-0801, Japan }
\author{J. E. Sonier}
\address{Department of Physics, Simon Fraser University, Burnaby, British Columbia, Canada V5A 156}
\author{A. N. Price, R. I. Miller, and R. F. Kiefl}
\address{TRIUMF and Department of Physics, University of British Columbia, Vancouver, British Columbia, Canada V6T 1Z1}
\author{M. Nohara, H. Suzuki and H. Takagi}
\address{Graduate School of Frontier Sciences, University of Tokyo, Bunkyo-ku, Tokyo 113-8656, Japan}
\date{\today}

\maketitle

\begin{abstract}
The magnetic field distribution in the vortex state of 
\yn\ has been probed by muon spin rotation (\msr).
The analysis based on the London model with nonlocal corrections shows 
that the vortex lattice has changed from hexagonal to square with 
increasing magnetic field $H$. At low fields the vortex core radius, 
$\rho_v (H)$, decreases with increasing $H$
much steeper than what is expected 
from the $\sqrt H$ behavior of the Sommerfeld constant $\gamma(H)$, 
strongly suggesting that the anomaly in $\gamma(H)$ primarily arises from the
quasiparticle excitations outside the vortex cores. 
\end{abstract}

]

The recent studies of the flux-line lattice (FLL) state in ordinary $s$-wave 
superconductors have revealed that the electronic 
structure of vortices is much more complicated than that of a 
simple array of rigid cylinders containing normal electrons. 
One of the unexpected phenomena within this conventional model 
is the non-linearity in the magnetic field dependence of the 
Sommerfeld constant $\gamma(H)$ (electronic specific heat coefficient) 
observed in CeRu$_2$\cite{Hedo:98}, NbSe$_2$\cite{Nohara:99}, and 
\yn\cite{Nohara:99}. According to the above simple model where 
the quasiparticle excitations are confined within the cores of vortices 
(with a radius $\xi$) in $s$-wave superconductors, one would expect 
that $\gamma(H)$ is proportional to the number of 
vortices per unit cell and thus to the applied magnetic 
field $H$. However, experiments have revealed that this is not 
the case for any of the above compounds.\cite{Hedo:98,Nohara:99} 
Instead, they find a field dependence like $\gamma(H) \propto \sqrt{H}$ 
which is expected for $d$-wave 
superconductors having more extended quasiparticle excitations along 
nodes in the energy gap. The recent study on the effect of doping in \yn\ 
and NbSe$_{2}$ indicates that the anomalous field dependence is 
observed only in the clean limit\cite{Nohara:99}, suggesting the
importance of nonlocal effects in understanding the field dependence of
$\gamma(H)$. Moreover, it has been reported that the vortex core radius
depends on applied magnetic field and shrinks at higher fields in 
NbSe$_{2}$ \cite{Sonier:97} and in CeRu$_{2}$ \cite{Kadono:01}.  
 
Another complication especially for borocarbides 
(RNi$_{2}$B$_{2}$C, R = rare earth) is that a square FLL 
is formed in some of these compounds 
at high magnetic fields, whereas a hexagonal FLL is realized at low 
fields. \cite{Yaron:96,Eskildsen:97,Paul:98,Eskildsen:98} 
This is not expected for the local model with isotropic
intervortex interactions and thereby suggests the importance
of considering electronic structure 
(or the Fermi surface) and the associated nonlocal corrections
in the specific compound.

We report on \msr\ measurements of the magnetic field dependence of the 
$\hat{a}$-$\hat{b}$ magnetic penetration depth $\lambda$, 
the effective vortex core radius $\rho_v$, and the apex angle of the 
FLL $\theta$ in single crystalline \yn. 
We demonstrate that  the proper reconstruction of 
the field profile with a square FLL is obtained from the \msr\ spectra 
only when the nonlocal corrections are considerred.\cite{Kogan:97}
The field dependence of $\lambda$ turned out to 
be linear over the entire magnetic field range of observation. 
More importantly, it was found that $\rho_v$ shrinks sharply with increasing 
magnetic field and levels off at higher fields. 
This shrinkage, however, is much steeper than that expected
for the case when the $\sqrt H$ behavior of $\gamma(H)$ is entirely
attributed to that of $\rho_v$, 
strongly suggesting that the anomaly in $\gamma(H)$ is mostly from the
quasiparticle excitations outside the vortex cores. 

The single crystal of \yn\ used in this experiment (residual resistivity 
ratio, or rrr $\simeq$ 37.4) had a 
surface area of $\sim$ 64 mm$^{2}$. The superconducting 
transition temperature $T_{c}$ and the upper critical field 
$H_{c2}$ ($T$ = 3 K) determined 
from resistivity and specific heat measurements were 15.4~K and 
7.0~T, respectively. \cite{Nohara:99} 
\msr\ experiments were performed on the M15 and M20 surface 
muon beamlines at TRIUMF. An experimental setup 
with high timing resolution was employed to measure the transverse field (TF-) 
\msr\ time spectra up to 5~T. The sample was mounted with its 
$\hat{c}$-axis parallel to the applied field and beam directions, while
the initial muon spin polarization was perpendicular to the 
applied field. The sample was field cooled at the measured 
magnetic fields to minimize disorder of the FLL due to flux pinning. 
Since the muons stop randomly on the length scale of the 
FLL, the muon spin precession signal provides 
a random sampling of the internal field distribution in 
the FLL state. 

\begin{figure}[htbp]
\begin{center}
\mbox{\epsfxsize=0.45\textwidth \epsfbox{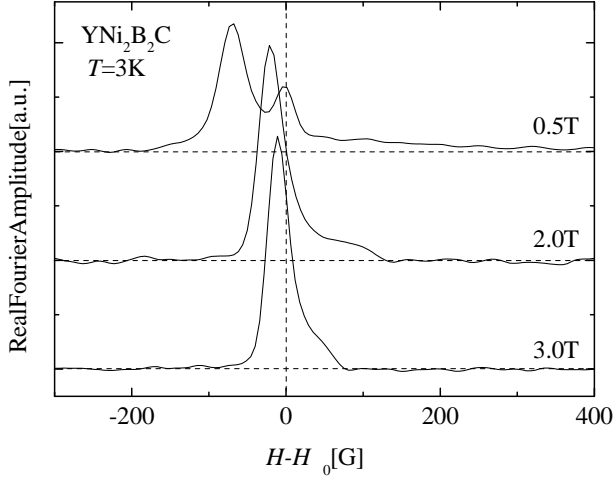}}
\end{center}
\caption{Fourier Transform of the \msr\ time spectra
in \yn\ at 3 K with a strong apodization (see text).}
\label{fig1}
\end{figure}

Figure \ref{fig1} shows the fast Fourier transforms (FFT) of the 
muon precession signal in \yn\ for different fields at $T \simeq$ 3.0 K 
with strong apodization \cite{Norton:76}. 
The real amplitude of the FFT corresponds to the internal magnetic field 
distribution in the FLL state convoluted with an additional damping 
to account for the weak nuclear dipolar fields, FLL disorder, and distortions 
originating from the finite time window and the reduced statisties at later 
times. \cite{Sonier:00} 
The high-field cutoff reflecting the magnetic field at the vortex core 
is clearly observed. The small peak near $H-H_0=0$ 
is the residual background generated by muons which missed the sample. 

In our preliminary analysis \cite{Ohishi:00}, it was revealed that the 
local London model with a square FLL fails to reproduce the observed 
\msr\/ spectra in \yn. 
More specifically, the apex angle $\theta$ of the FLL
gradually increases from 60$^\circ$ with increasing field, but it
levels off over the field range above 0.5 T with $\theta\simeq 75^\circ$ 
(see Fig. \ref{fig3}(c)) where the square FLL is established by other measurements 
(i.e., $\theta\simeq 90^\circ$)\cite{Paul:98,Yethiraj:97,Yethiraj:98,Sakata:00}. 
Thus, the result in Ref.\onlinecite{Ohishi:00} was obtained with $\theta\simeq 75^\circ$
for $H>0.5$ T.
As it is demonstrated below,
we have found that this problem is alleviated 
by taking account of the nonlocal corrections \cite{Kogan:97}. 
The local magnetic field at any point in the $\hat{a}$-$\hat{b}$ plane is 
\begin{equation}
H({\bf r})=\overline{H}_{0}\sum_{\bf K}\frac{e^{-i{\bf K}\cdot{\bf r}}e^{-K^{2
}\xi_{v}^{2}}}{1+K^{2}\lambda^{2}+\lambda^{4}(0.0705CK^{4}+0.675Ck_{x}^{2}k_{y}^{2})}, 
\label{eq1}
\end{equation}
where {\bf K} is the reciprocal lattice vector, 
\begin{figure}[htbp]
\begin{center}
\mbox{\epsfxsize=0.4\textwidth \epsfbox{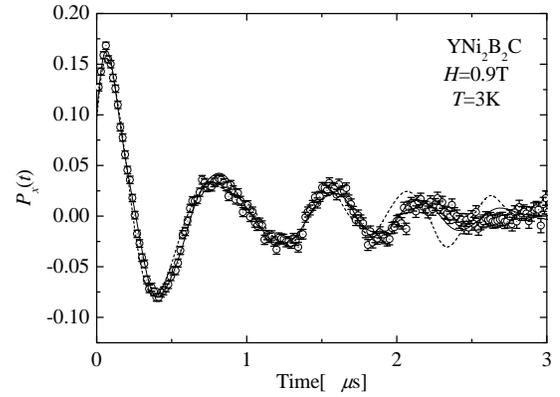}}
\end{center}
\caption{The muon precession signal 
$P_x(t)$ in \yn\/ at $H$ = 0.9 T, displayed in a rotating-reference-frame
frequancy of $\sim$2 MHz. For the solid/dashed curves, see the text.}
\label{time}
\end{figure}
\begin{eqnarray}
{\bf K}&=&l{\bf k}_{x}+m{\bf k}_{y}, \quad 
(l,m=0, \pm1, \pm2, ...) \label{eq3}\\
{\bf k}_{x}&=&\frac{2\pi}{a\sin\theta}
\left(-\cos\frac{\theta}{2}\hat{x}+\sin\frac{\theta}{2}\hat{y}\right) \label{eq4}\\
{\bf k}_{y}&=&\frac{2\pi}{a\sin\theta}
\left(\cos\frac{\theta}{2}\hat{x}+\sin\frac{\theta}{2}\hat{y}\right),\label{eq5}
\end{eqnarray}
with $\hat{x}$ and $\hat{y}$ being the plane of precession, $a$ the 
FLL parameter, $\theta$ the apex angle of the FLL, 
$\overline{H}_{0}$ the average magnetic field, $\lambda$ the 
magnetic penetration depth, and $\xi_{v}$ being the cutoff parameter. 
The above reciprocal lattice vectors correspond
to the case where the diagonal direction of the FLL 
(= ${\bf u}+{\bf v}$ = 2$\sin{\frac{\theta}{2}}\hat{x}$) 
is along the $\langle$100$\rangle$ direction of the crystal axis.
The anisotropic parameter $C$ is determined by the band structure, in which
$C$ scales with $\lambda$ as $C=C_0/\lambda^2$.\cite{Kogan:97}
The coefficients for $C$ in Eq.(\ref{eq1}) were adopted from the theoretical
estimation for LuNi$_2$B$_2$C.\cite{Kogan:97}
The local London model is obtained by putting $C=0$. 

In addition to the nonlocal corrections, we have developed 
a program to analyze the \msr\ spectra in 
the {\it time} domain to eliminate the uncertainty in the
estimation of statistical errors associated with fitting 
the FFT spectra. The theoretical time evolution of the muon spin polarization 
was generated by assuming the field profile of Eq.~(\ref{eq1}) \cite{Sonier:00},
\begin{equation}
P_x(t)+iP_y(t)=\int \frac{d{\bf r}}{dH({\bf r})}\exp(i\gamma_\mu H({\bf r})t)dH
\label{complex}
\end{equation}
($\gamma_\mu$ is the muon gyromagnetic ratio) 
and compared with the time spectra by the chi-square ($\chi^2$) minimization technique.
Considering the results of small angle neutron scattering (SANS) 
\cite{Paul:98,Yethiraj:97,Yethiraj:98} and scanning tunneling 
microscopy/spectroscopy (STM/STS) \cite{Sakata:00}, 
the apex angle $\theta$ was fixed to 90$^{\circ}$ for $H \ge$ 0.4 T while
it was treated as a fitting parameter for $H<$ 0.4 T. 

\begin{figure}[htbp]
\begin{center}
\mbox{\epsfxsize=0.28\textwidth \epsfbox{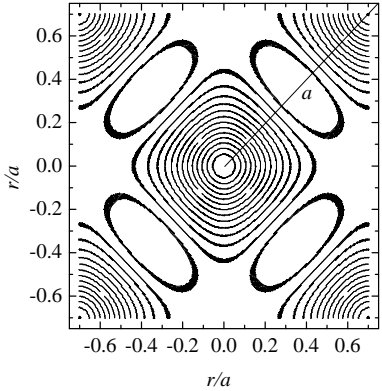}}
\end{center}
\caption{The contour map of a flux-line at $H$ = 0.9 T in real space, where 
the unit cell length $a$ of FLL is 471.9 \AA.}
\label{fig2}
\end{figure}

\begin{figure}[htbp]
\begin{center}
\mbox{\epsfxsize=0.45\textwidth \epsfbox{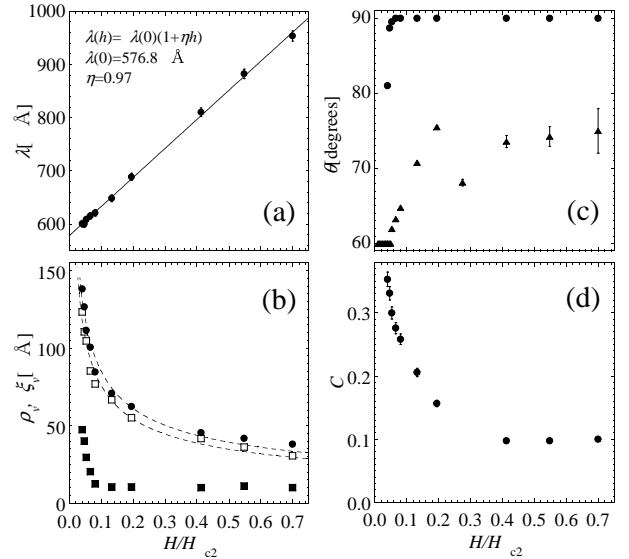}}
\end{center}
\caption{The $H$ dependence of (a) $\lambda$, (b) $\rho_v$ determined by \msr\ where 
$\rho_{v\langle100\rangle}$ is shown by open squares, 
$\rho_{v\langle110\rangle}$ by solid circles and $\xi_{v}$ by solid squares, 
(c) $\theta$ by circles and the one deduced from the local London model \cite{Ohishi:00} by triangles, 
and (d) $C$ in the FFL state of \yn\/ at 3 K. 
The dashed curves in (b) are described in the text.}
\label{fig3}
\end{figure}

A typical example of the \msr\ time spectra measured in \yn\ under a magnetic field of 0.9 T 
is shown in Fig. \ref{time}, where the solid curve is a fit by the nonlocal London model 
while the dashed curve is by the local model with the apex angle fixed to 90$^\circ$. 
The value of deduced $\chi^2$ for the nonlocal model is more than three
times smaller than that for the local model, indicating that the nonlocal model provides
much better description of the data. The rate of additional Gaussian relaxation due to 
trivial sources (nuclear dipolar fields, vortex pinnning, etc.) is about 0.34 $\mu$s$^{-1}$ 
at 0.9 T and it tends to be idependent of the field. 
Figure \ref{fig2} shows the contour plot of $H({\bf r})$ around 
a vortex at $H$ = 0.9~T reproduced from \msr\ data, 
where the $\langle$100$\rangle$ axis of the crystal is along the 
horizontal direction. The fourfold symmetry due to the nonlocal 
corrections in Eq.(\ref{eq1}) is clearly observed. We note that
there are two possible orientations of the FLL configuration in Eq.(\ref{eq1}),
where the diagonal direction ${\bf u}+{\bf v}$ is parallel with either the 
$\langle$100$\rangle$ or the $\langle$110$\rangle$ crystalline axis.
We have found that the field distribution with ${\bf u}+{\bf v}$ parallel
with $\langle$110$\rangle$ does not reproduce our data with 
any combination of parameters.
This is perfectly in line with the results of other experiments, as well as 
the theoretical calculation which yields a lower free energy for 
${\bf u}+{\bf v}$ parallel with $\langle$100$\rangle$.\cite{Kogan:97}

The physical parameters $\lambda$, $\xi_{v}$, $\theta$ and $C$ versus 
normalized external field at 3 K  are shown in Fig. \ref{fig3}. 
We treated $C$ as a fitting parameter because its value in \yn\ is unknown. 
The 
$\lambda$ in \yn\ clearly 
exhibits a {\it linear} $H$-dependence. A fit to the relation 
$\lambda(h) = \lambda(0)(1+\eta\cdot h), \ (h=H/H_{c2})$
provides a dimensionless parameter $\eta$ that represents the strength 
of the pair-breaking effect. We obtain $\eta$ = 0.97 (with $\lambda(0)$ 
= 567.8 \AA) which is slightly smaller than that in NbSe$_{2}$ (i.e., 
$\eta$ = 1.61 at 0.33$T_{c}$ \cite{Sonier:97}). 
The cutoff parameter $\xi_v$ (Fig. \ref{fig3}(b), solid squares) shows 
a steep decrease with increasing $H$ and subsequently levels off at 
$h\equiv H/H_{c2} >0.1$ ($H >$ 0.7 T). In our preliminary analysis \cite{Ohishi:00}, 
we interpreted this cutoff parameter as $\rho_v$
(see eq.~(2) in Ref.\onlinecite{Ohishi:00}). 
In the field region $h<$ 0.06 where $\theta$ was set as
a free parameter, $\theta$ gradually decreases with decreasing 
field, indicating that the FLL transforms into a nearly hexagonal lattice. 
However, $\theta$ does not reach 60$^{\circ}$ in the lowest magnetic 
field. The anisotropy $C$ decreases with increasing $H$,
where it exhibits little correlation with $\theta$. 
While the value at lower field is close to the theoretical 
estimation ($\sim$ 0.22 at 0.05 T in LuNi$_2$B$_2$C \cite{Kogan:97}),
we found that $C\lambda^2$ tends to decrease with increasing field.

The field profile in Fig.~\ref{fig2} implies that there is an anisotropy between the 
$\langle$100$\rangle$ and $\langle$110$\rangle$ directions in the effective 
length scales ($\lambda$ and $\rho_v$), whereas the model 
parameters in Eq.(\ref{eq1}) represent mean values. 
Moreover, special precaution must be taken to
interpret the parameters in Eq.(\ref{eq1}) upon the introduction of nonlocal
corrections involving the higher order terms of $K$,
where the definition of these length scales are modified from
those found in the previous analysis with local London models, making
it unsuitable to compare directly.
In order to evaluate $\rho_v$ including the effect of anisotropy,
we calculated the supercurrent density $J(r)$ from the 
deduced $H(r)$ using Maxwell's relation 
$J$({\bf r})~=~$\mid\nabla \times$ {\bf H}({\bf r})$\mid$.
The radius $\rho_{v}$ was then defined
as the distance from the vortex center for which $J(r)$ 
reaches its maximum value. 
The estimated values are $\rho_{v\langle100\rangle}$~=~66.7\AA\/ 
and $\rho_{v\langle110\rangle}$~=~70.8\AA, yielding the ratio 
$\rho_{v\langle100\rangle}/\rho_{v\langle110\rangle}$~=~0.942. 

The field dependence of $\rho_{v}$ is shown in Fig. \ref{fig3}(b). 
The values of $\rho_v$ are systematically larger than $\xi_v$, suggesting that 
it may not be appropriate to interpret the cutoff parameter as the vortex 
core radius,
whereas we assumed that $\rho_v=\xi_v$ in the previous analysis\cite{Ohishi:00}.
We stress that the core radius can be obtained directly 
from the field profile $H({\bf r})$ deduced from the \msr\ data, independent 
of the details of the FLL model used. \cite{Sonier:00}
Having said this, the field dependence of $\rho_v$ is qualitativly similar to 
that of $\xi_v$, showing a steep decrease with increasing field in
the field range $H/H_{c2} <$ 0.15.  

Recent calculations for $s$-wave superconductors based on the quasiclassical 
Eilenberger equations predicts a shrinkage of 
$\rho_{v}$ due to vortex-vortex interactions. \cite{Ichioka:99} 
The quasiparticle density 
of states (DOS) $N(H)$ is proportional to $H^{\beta}$ with $\beta$ = 0.67 
at $T$ = 0 in their prediction. 
Provided that all the DOS comes from inside the vortex cores, 
we would expect
\begin{equation}
N(H)=N_{\rm core}(H) \propto \pi\rho_{v}^{2} \cdot H \propto H^{\beta}
\label{qdos}
\end{equation}
where the factor $H$ arises from the number of vortices per unit area, and 
$\rho_{v} \propto H^{(\beta-1)/2}$.
Fitting the field dependence of $\rho_{v\langle100\rangle}$ and 
$\rho_{v\langle110\rangle}$ in 
Fig. \ref{fig3}(b) to the relation $\rho_{v}=\rho_{0}h^{(\beta-1)/2}$
yields $\beta_{\langle100\rangle}$ = 0.026, 
$\rho_{0 \langle100\rangle}$ = 24.7\AA\/ and 
$\beta_{\langle110\rangle}$ = 0.014, $\rho_{0 \langle110\rangle}$ = 27.1\AA, 
while the field dependence of $\gamma(H)$ yields 
$\beta=0.430\equiv\beta_{\rm SH}$ \cite{Nohara:99}. 
The considerably smaller values for
$\beta_{\langle100\rangle}$ and $\beta_{\langle110\rangle}$ 
compared with $\beta_{\rm SH}$ or the theoretical prediction
strongly suggests that the origin of the $\sqrt H$ behavior of $\gamma(H)$ is 
related to the quasiparticle excitations outside the 
vortex cores. This is in marked contrast with the case of CeRu$_2$ where 
$\beta\simeq\beta_{\rm SH}$, indicating that the DOS is mostly 
attributed to the quasiparticles within the vortex cores. \cite{Kadono:01}
The existence of delocalized quasiparticle excitations is further suggested 
by the fact that de Haas-van Alphen effect has been clearly observed 
in the mixed state of \yn, where the cyclotron radius is much larger 
than the coherence length $\xi$. \cite{Terashima:97} 
Surface impedance $Z_{s}$ measurements also indicate delocalized 
quasiparticles outside the vortex cores. \cite{Izawa:01} 
The magnetic field dependence of $N(H)$ inside the cores estimated by $Z_{s}$ 
is proportional to $H$, except at very low field. 
These results are consistent with our conclusion that the localized 
quasiparticles within the vortex cores (determined by $\rho_v$) contribute 
little to the $\sqrt H$ behavior of the Sommerfeld constant, at least for $h >$ 0.15. 
Here, we note that the agreement between $\beta$ and $\beta_{\rm SH}$
is improved by assuming that $N(H)\propto \rho_v\cdot H$ 
instead of Eq.~(\ref{qdos}) \cite{Sonier:99}, 
although the microscopic origin of this linear relation is not
obvious at this stage. In any case, the small $\beta$ and associated 
steeper field dependence of $\rho_v$ at lower fields might 
be partly explained by the multiband effect, where the electronic structure is 
effectively described by a two-band model\cite{Shulga:98}.
The BCS coherence length $\xi_0=\hbar v_F/\pi\Delta_0$ (where 
$\rho_v\le 0.6\xi_0$ \cite{Ichioka:99}) estimated
from the Fermi velocity $v_F$ and the energy gap $\Delta_0$ in \yn, is 
60 to 120 \AA\ for one group and 370 \AA\ for another branch, suggesting that
$\rho_v(H\rightarrow0)$ is controlled by the larger value of $\xi_0$.
We also note a possible connection to the anisotropic energy gap in 
\yn\/ reported
by photoemission spectroscopy \cite{Yokoya:00}, where $\xi_0$ is
scaled by the magnitude of $\Delta_0$.

Finally, we discuss the apex angle $\theta$ at lower fields where the
deviation from a hexagonal lattice is expected ($h < 0.04$). 
We found that the agreement between the measured field distribution 
and calculations based on the
present model becomes far from satisfactory in the field range at $h <$ 0.04 ($H <$ 0.3 T). 
This is probably due to the presence of the deep minima along the 
$\langle110\rangle$ direction in Fig.~\ref{fig2}, which persists irrespective 
of the apex angle (Note that the 
square shaped field distribution is independent of $\theta$, as is evident
in Eq.(\ref{eq1})). 
The poor agreement strongly suggests that Eq.(\ref{eq1})
gives the true ground state only for the case of a square FLL, while the more
isotropic distribution would be realized at lower fields 
as the $\theta \simeq 60^{\circ}$ well reproduced by the local London model \cite{Ohishi:00}. 
Thus, a more refined model is needed to reproduce the complete evolution of the FLL with 
field. We also point out the possibility that FLL domains present through the 
hexagonal-to-square transition (like in LuNi$_2$B$_2$C \cite{Vinnikov:00}) 
play an important role.

In summary, we found that $\rho_v$ shrinks steeply with increasing field while 
$\lambda$ depends linearly on the magnetic field,
strongly suggesting the presence of excess quasiparticles outside
the vortex cores at higher fields.
These results indicate the need to reconsider the 
conventional picture of a rigid normal-electron core by taking into 
account the vortex-vortex interactions mediated by delocalized quasiparticles.

We thank the TRIUMF \msr\ staff for technical support. 
This work was partially supported by a JSPS Research Fellowships for Young Scientists, 
Japan, by a Grant-in-Aid for Science Research 
on Priority Areas from the Ministry of Education, Culture, Sports, Science, 
and Technology, Japan and also by a Grant from the CREST, JST, Japan.

\end{document}